\documentstyle[epsfig]{aipproc}

\def\Dunits{$\times10^{28}$ cm$^2$s$^{-1}$}
\begin{document}

   \title{Modelling cosmic rays \\ and gamma rays in the Galaxy}
   \author{Andrew W. Strong$^\star$ and Igor V.~Moskalenko$^{\star\dagger}$}
   \address{
      $^\star$Max-Planck-Institut f\"ur extraterrestrische Physik,
      D-85740 Garching, Germany \\
      $^\dagger$Institute for Nuclear Physics,
      Moscow State University, 119 899 Moscow, Russia }
   \maketitle

\begin{abstract}
An extensive program for the calculation of galactic cosmic-ray
propagation has been developed.  This is a continuation of the work
described in \cite{StrongYoussefi95}.  The main motivation for
developing this code \cite{StrongMoskalenko97} is the prediction of
diffuse Galactic gamma rays for comparison with data from the {\sc
cgro} instruments {\sc egret}, {\sc comptel}, and {\sc osse}.  The basic
spatial propagation mechanisms are (momentum-dependent) diffusion,
convection, while in momentum space energy loss and diffusive
reacceleration are treated.  Primary and secondary nucleons, primary
and secondary electrons, and secondary positrons are included.
Fragmentation and energy losses are computed using realistic
distributions for the interstellar gas and radiation fields.

\end{abstract}

\section*{Introduction}
We are developing a model which aims to reproduce self-consistently
observational data of many kinds related to cosmic-ray origin and
propagation: direct measurements of nuclei, electrons and positrons,
gamma rays, and synchrotron radiation. These data provide many
independent constraints on any model and our approach is able to take
advantage of this since it must be consistent with all types of
observation.  We emphasize also the use of realistic astrophysical
input (e.g., for the gas distribution) as well as theoretical
developments (e.g., reacceleration).  The code is sufficiently general
that new physical effects can be introduced as required.  The basic
procedure is first to obtain a set of propagation parameters which
reproduce the cosmic ray B/C ratio, and the spectrum of secondary
positrons;  the same propagation conditions are then applied to primary
electrons. Gamma-ray and synchrotron emission are then evaluated.
Models both with and without reacceleration are considered.  The models
are three dimensional with cylindrical symmetry in the Galaxy, the
basic coordinates being $(R,z,p)$ where $R$ is Galactocentric radius,
$z$ is the distance from the Galactic plane and $p$ is the total
particle momentum.  The numerical solution of the transport equation is
based on a Crank-Nicholson implicit second-order scheme. In the models
the  propagation region is bounded by $z=z_h$ beyond which free escape
is assumed.  A value $z_h=3$ kpc  has been adopted since this is within
the  range which is consistent with studies of $^{10}$Be/Be and
synchrotron radiation. For a given $z_h$ the diffusion coefficient as a
function of momentum is determined by B/C for the case of no
reacceleration; with reacceleration on the other hand it is the
reacceleration strength (related to the Alfv\'en speed $v_A$) which is
determined by B/C.  Reacceleration provides a natural mechanism to
reproduce the B/C ratio without an ad-hoc form for the diffusion
coefficient \cite{HeinbachSimon95,SeoPtuskin94}.  The spatial diffusion
coefficient for the case {\it without} reacceleration is $D=\beta D_0$
below rigidity $\rho_0$, $\beta D_0(\rho/\rho_0)^\delta$ above rigidity
$\rho_0$, where $\beta$ is the particle speed.  The spatial diffusion
coefficient {\it with} reacceleration assumes a Kolmogorov spectrum of
weak MHD turbulence so $D=\beta D_0(\rho/\rho_0)^\delta$ with
$\delta=1/3$ for all rigidities.  For this case  the momentum-space
diffusion coefficient is related to the spatial one
\cite{SeoPtuskin94}.  The injection spectrum of nucleons is assumed to
be a power law in momentum.  The interstellar hydrogen distribution
uses HI and CO surveys and information on the ionized component; the
Helium fraction of the gas is taken as 0.11 by number.  The
interstellar radiation field for inverse Compton losses is based on
stellar population models and {\sc iras} and {\sc cobe} data, plus the
cosmic microwave background.  Energy losses for electrons by
ionization, Coulomb, bremsstrahlung, inverse Compton and synchrotron
are included, and for nucleons by ionization and Coulomb interactions.
The distribution of cosmic-ray sources is chosen to reproduce the
cosmic-ray distribution determined by analysis of {\sc egret} gamma-ray
data \cite{StrongMattox96}.  The bremsstrahlung and inverse Compton
gamma rays are computed self-consistently from the gas and radiation
fields used for the propagation. The $\pi^0$-decay gamma rays are
calculated explicitly from the proton and Helium spectra using
\cite{Dermer86}. The secondary nucleon and secondary $e^\pm$ source
functions are computed from the propagated primary distribution and the
gas distribution, and the anisotropic distributions of $e^\pm$ in the
$\mu^\pm$ system was taken into account \cite{MoskalenkoStrong98}.

\section*{Illustrative results}
Some results obtained are shown in the Figures.  The energy dependence
of the B/C ratio, and local {\it proton and Helium spectra} are shown
in Figs.~\ref{fig1} and \ref{fig2}.  The spectrum of {\it primary
electrons} is shown in Fig.~\ref{fig6}.  The adopted electron injection
spectrum has a power law index --2.1 up to 10 GeV; this is chosen using
the constraints from synchrotron and from gamma rays.  The electron
spectrum is consistent with the direct measurements around 10 GeV where
solar modulation is small and it also satisfies the constraints from
${e^+}\over {e^-+\,e^+}$.  Above 10 GeV a break is required in the
injection spectrum to at least --2.4 for agreement with direct
measurements (which may however not be necessary if local sources
dominate the directly measured high-energy electron spectrum).  The
synchrotron spectrum at high Galactic latitudes (Fig.~\ref{fig3}) is
important since its shape constrains the shape of the 1--10 GeV
electron spectrum.  An injection index --2.1  (without reacceleration)
is the steepest which is allowed by the radio data over the range 38 to
1420 MHz. As illustrated, an index --2.4  as often used (e.g.,
\cite{Strong96}) gives a synchrotron spectrum which is too steep.

The modelled {\it gamma-ray} spectrum for the inner Galaxy, shown
here for the case of no reacceleration (Fig.~\ref{fig4}), fits well the
{\sc comptel} \cite{Strong97} and {\sc egret} \cite{StrongMattox96}
data between 1 MeV and 1 GeV beyond which there is the well-known
excess not accounted for by $\pi^0$-decay with the standard nucleon
spectrum \cite{Hunter97,Mori97}.  The electron spectrum used here is
increased by a factor 2 over that shown in Fig.~\ref{fig6}; this is
required to reproduce well the observed gamma intensities. It is still
within the range allowed by the positron fraction (see below), and the
shape accords well with the flatter electron injection spectrum
required by synchrotron data (Fig.~\ref{fig3}).  Inverse Compton is
dominant below 10 MeV, bremsstrahlung becomes important for 3--200
MeV.  The lower bremsstrahlung combined with  $\pi^0$-decay leads to a
good fit to the flat spectrum observed in this range, in contrast to
previous attempts to model the spectrum with a steeper bremsstrahlung
spectrum \cite{Strong96}.  The 10--30 MeV $\gamma$-ray longitude
profile (Fig.~\ref{fig5}) at low latitudes from this model can be
compared with that from {\sc comptel} \cite{Strong97}.  The $e^+$ {\it
spectrum} (Fig.~\ref{fig6}) and the  {\it positron fraction}
(Fig.~\ref{fig7}) agree well with the most recent data compilation for
0.1--10 GeV \cite{MoskalenkoStrong98}.

This study indicates that it is possible to construct a model
satisfying a wide range of observational constraints and provides a
basis for future developments.

\smallskip
\footnotesize More details can be found on 
{\it http://www.gamma.mpe--garching.mpg.de/$\sim$aws/aws.html}

\newpage

\def\figheight{55mm}
\def\figwidth{80mm}
\begin{figure}[t!]
   \begin{picture}(140,49)(1,0)
      \put(-4,-2){ \makebox(80,55)[l]{ \psfig{file=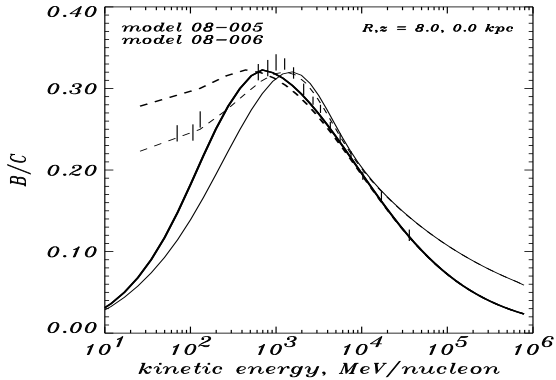,%
         height=\figheight,width=\figwidth,clip=}}}
      \put(80,0){ \begin{minipage}[b]{58mm}
         \caption[]{%
The energy dependence of the B/C ratio can be reproduced with $D_0 =
2.0$\Dunits, $\delta=0.6$, $\rho_0= 3$ GV/c without reacceleration
(thick line) and $D_0 = 4.2$\Dunits, $v_A=20$ km s$^{-1}$ with
reacceleration (thin line).  Dashed lines are modulated to 500 MV. Data
from \cite{Webber96}.}
         \label{fig1}
      \end{minipage}}
   \end{picture}
\end{figure}

\def\figheight{55mm}
\def\figwidth{70mm}
\begin{figure}[t!]
   \begin{picture}(140,59)(5,0)
      \put(0,0){ \makebox(70,57)[tl]{ \psfig{file=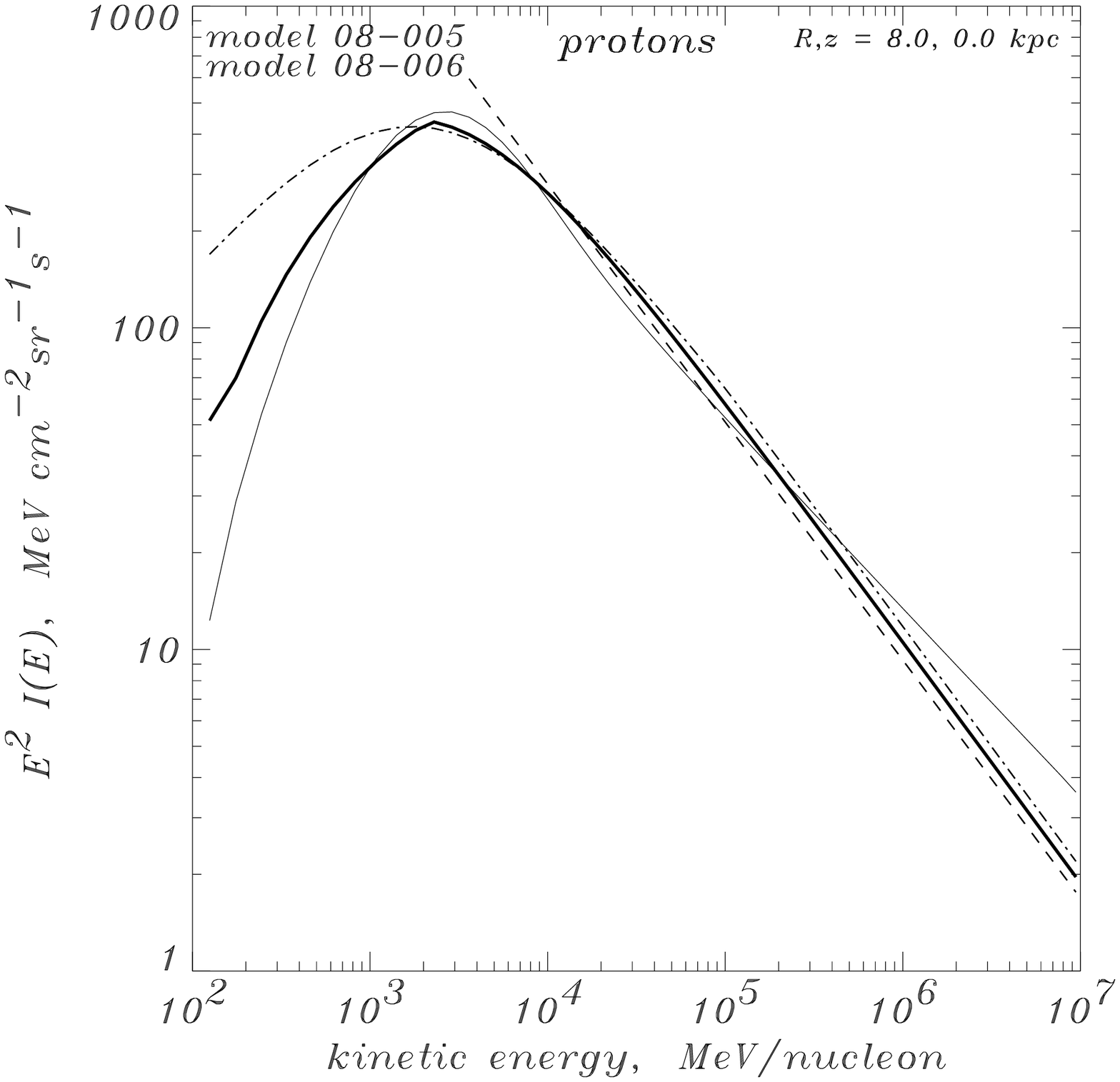,%
         height=\figheight,width=\figwidth,clip=}}}
      \put(70,0){ \makebox(70,57)[tl]{ \psfig{file=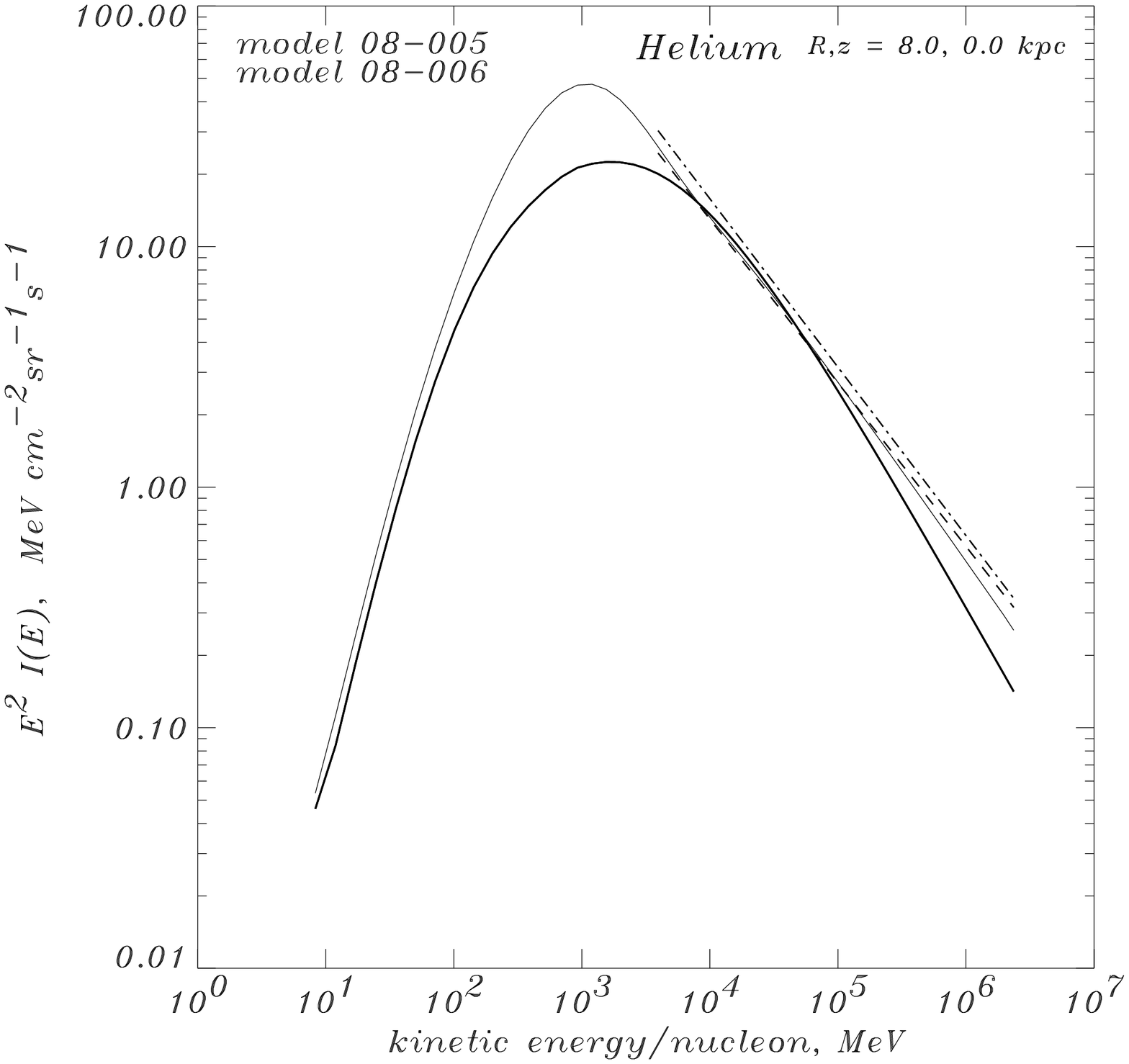,%
         height=\figheight,width=\figwidth,clip=}}}
   \end{picture}
   \caption[]{%
{\it Left panel:} the local proton spectrum for injection index 2.15
(thin solid line), 2.25 (thick solid line) without and with
reacceleration respectively, compared with the measured `interstellar'
spectrum (dashed \cite{Seo91} and dashed-dot \cite{Mori97} lines).
{\it Right panel:}  the Helium spectrum with injection index 2.25 (thin
solid line), 2.45 (thick solid line)  without and with reacceleration
respectively, compared with the measured `interstellar' spectrum
(dashed \cite{Seo91} and  dashed-dot \cite{Engelmann90} lines).}
   \label{fig2}
\end{figure}

\def\figheight{55mm}
\def\figwidth{72mm}
\begin{figure}[t!]
   \begin{picture}(140,70)(1,0)
      \put(-6,13){ \makebox(70,57)[l]{ \psfig{file=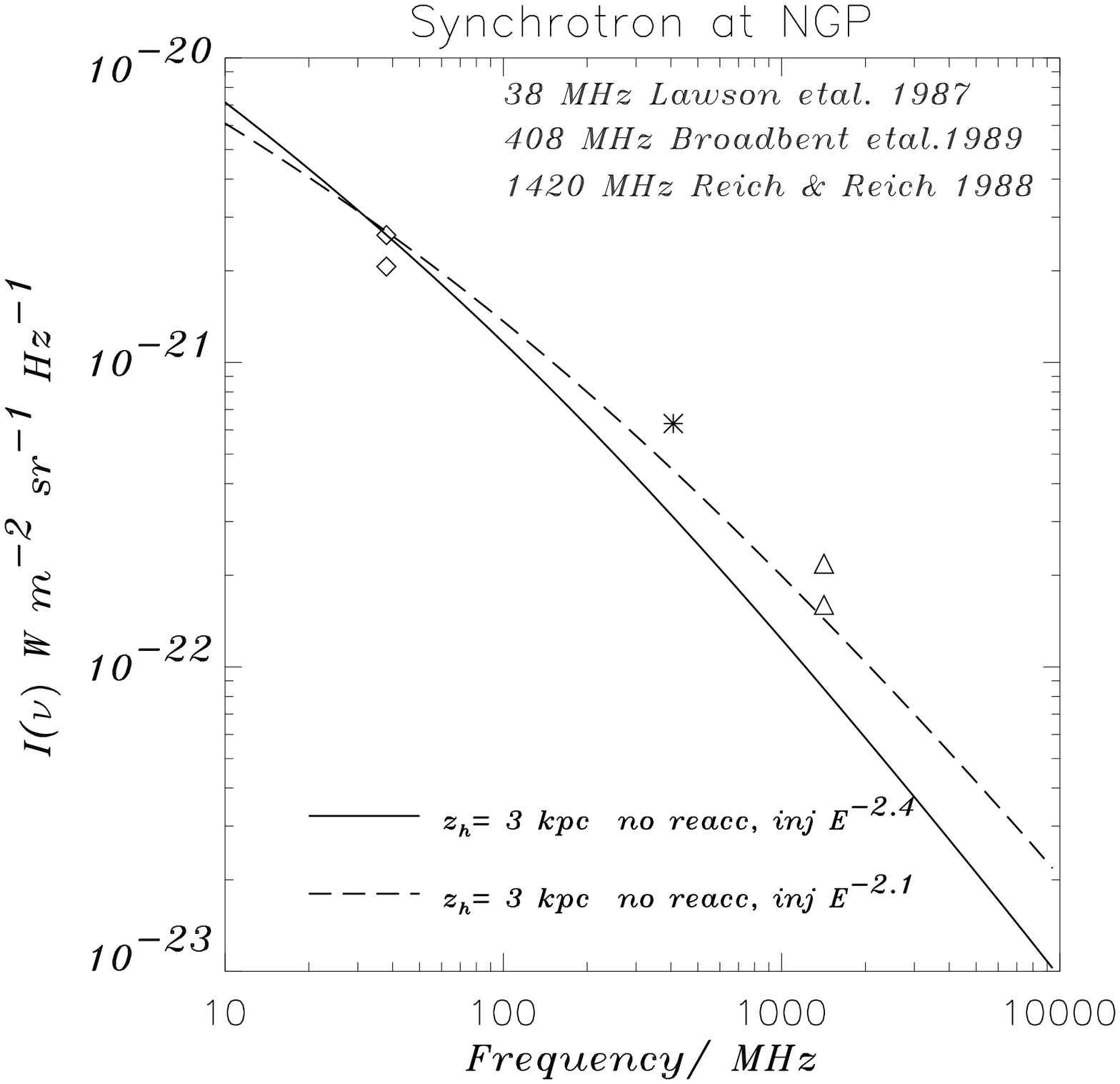,%
         height=\figheight,width=\figwidth,clip=}}}
      \put(0,13){ \begin{minipage}[t]{71mm} 
         \caption[]{%
The synchrotron spectrum at the NGP, and predictions for electron
injection indices --2.1 (dashed line) and --2.4 (solid line).}
         \label{fig3}
      \end{minipage}}

      \put(63,13){ \makebox(70,57)[l]{ \psfig{file=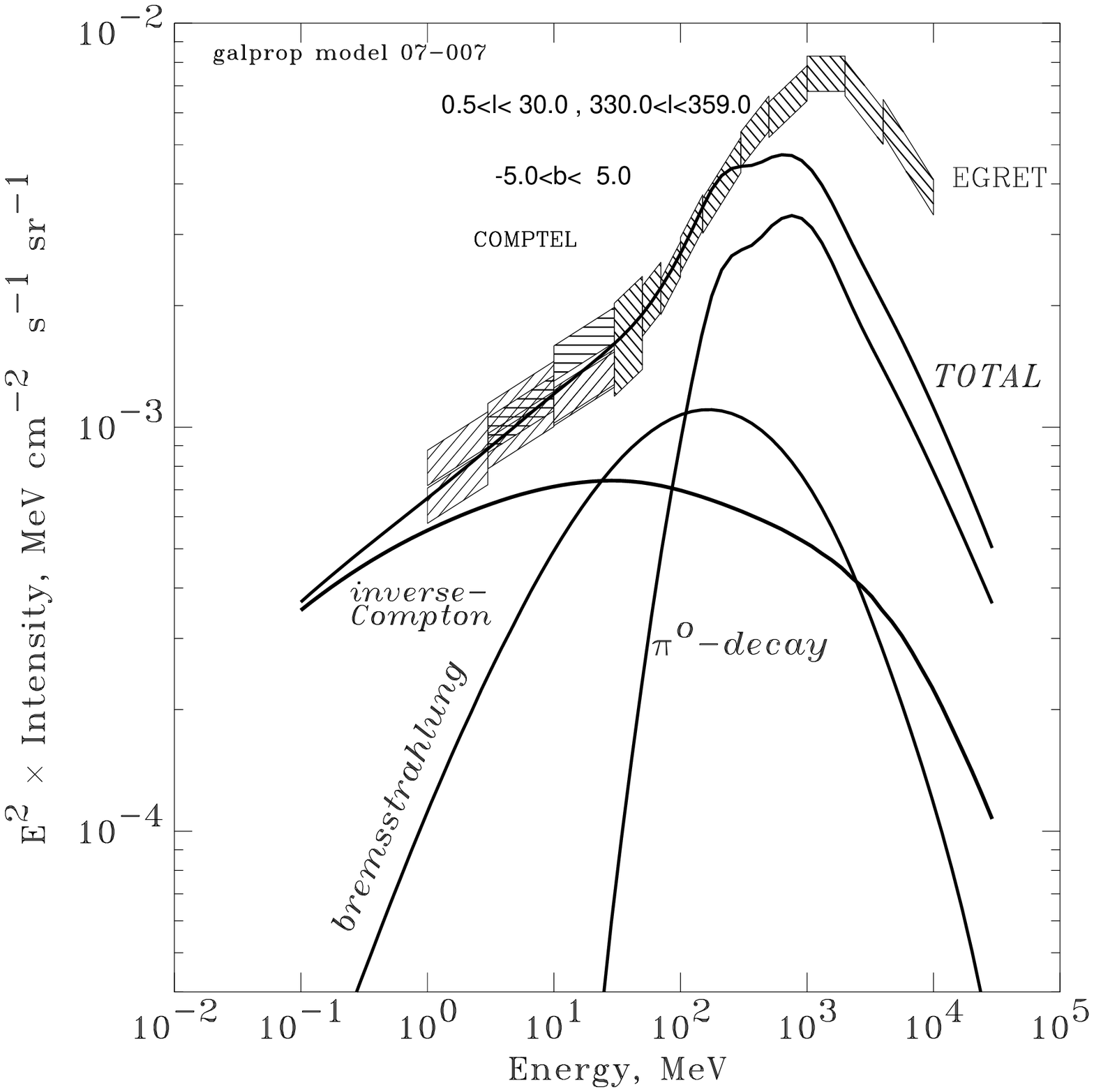,%
         height=\figheight,width=\figwidth,clip=}}}
      \put(75,13){ \begin{minipage}[t]{64mm}
         \caption[]{%
The $\gamma$-ray spectrum for the inner Galaxy, $330^\circ<l<30^\circ,
|b|<5^\circ$. {\sc egret} data \cite{StrongMattox96}, {\sc comptel}
data \cite{Strong97}.}
         \label{fig4}
      \end{minipage}}
   \end{picture}
\end{figure}

\def\figheight{59mm}
\def\figwidth{120mm}
\begin{figure}[t!]
   \begin{picture}(140,59)(0,0)
       \put(0,0){ \makebox(140,59)[l]{ \psfig{file=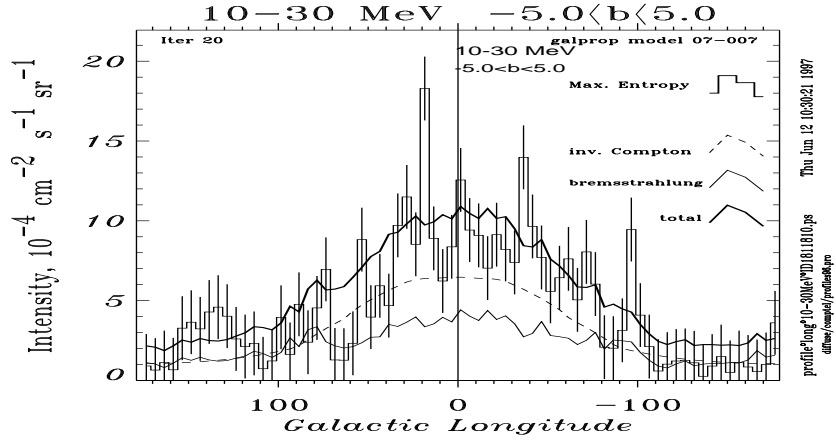,%
          height=\figheight,width=\figwidth,clip=}}}
   \end{picture}
   \caption[]{%
Longitude distribution of gamma rays in energy range 10-30 MeV.
Histogram: {\sc comptel} \cite{Strong97}, dashed: inverse Compton, thin
line:  bremsstrahlung, thick line: total.}
   \label{fig5}
\end{figure}

\def\figheight{90mm}
\def\figwidth{83mm}
\begin{figure}[t!]
   \begin{picture}(140,60)(7,-3)
      \put(-2,0){ \makebox(70,60)[tl]{ \psfig{file=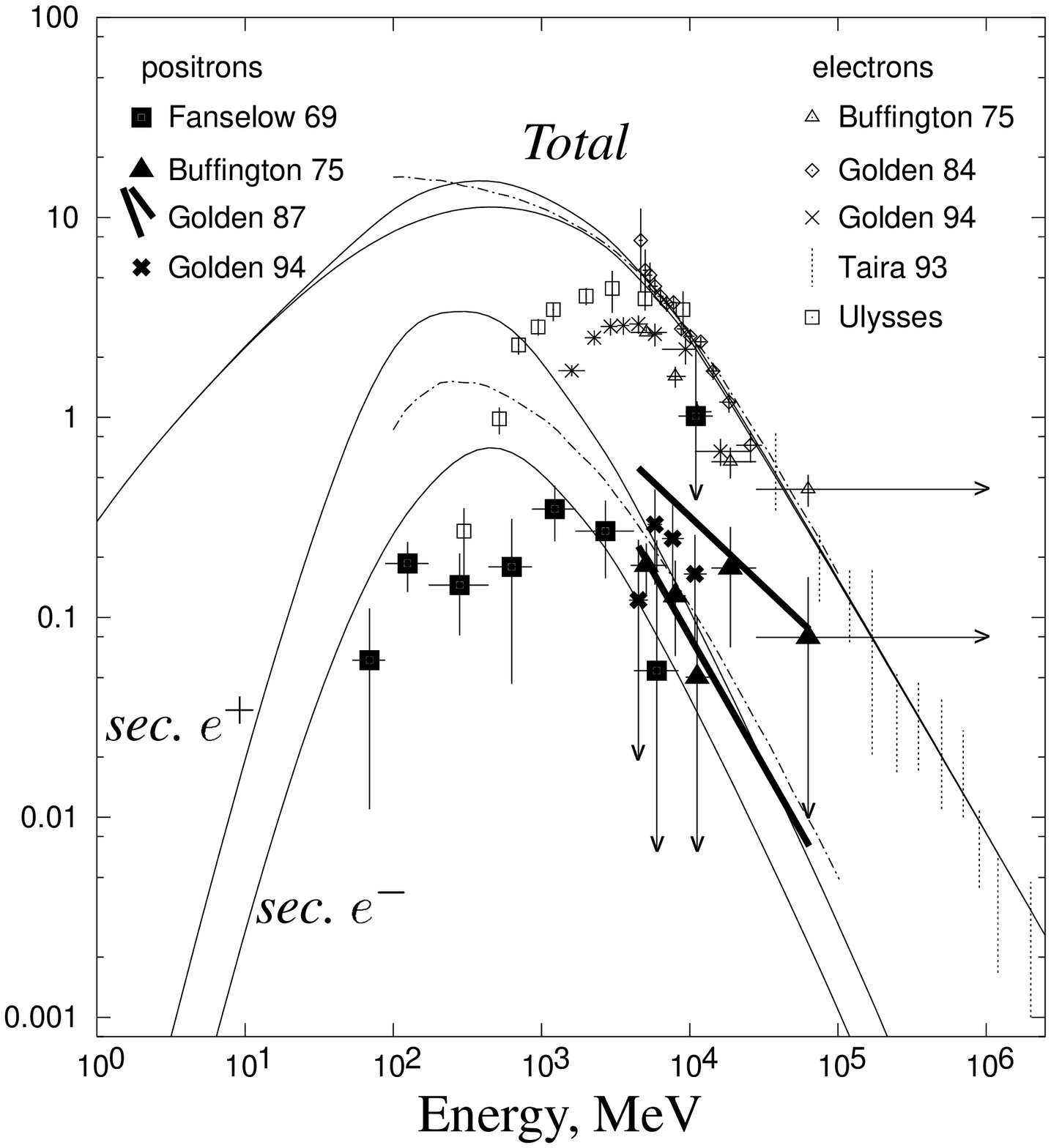,%
         height=\figheight,width=\figwidth,clip=}}}
      \put(69,0){\makebox(70,60)[tl]{\psfig{file=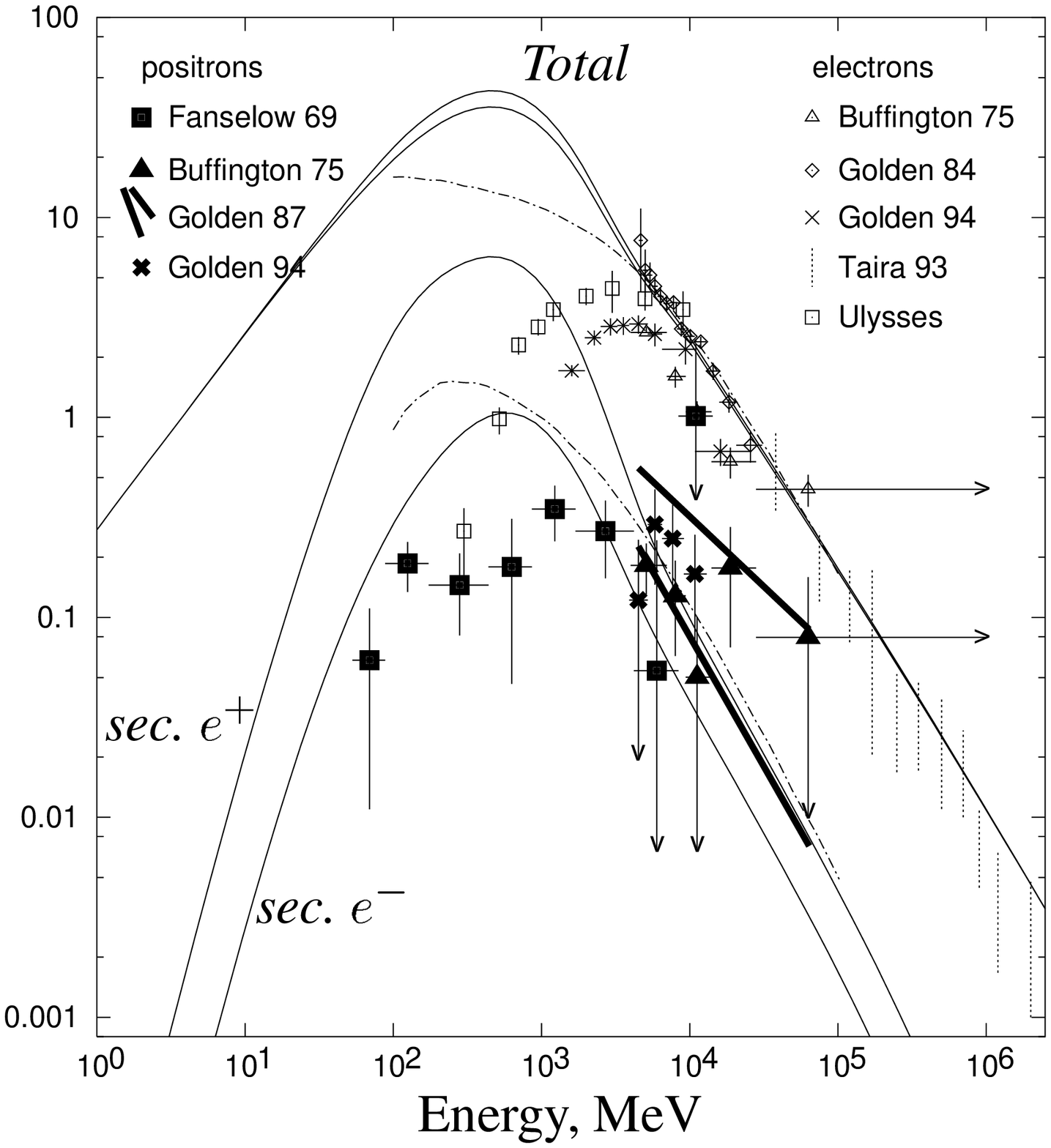,%
         height=\figheight,width=\figwidth,clip=}}}
   \end{picture}
   \caption[]{%
Spectra of secondary $e^\pm$, and of primary electrons. Full lines: our
model with no reacceleration (left) and with reacceleration (right).
Dash-dotted lines by Protheroe \cite{Protheroe82}: lower is his
leaky-box prediction for $e^+$, upper is his adopted electron
spectrum.}
    \label{fig6}
\end{figure}

\def\figheight{100mm}
\def\figwidth{80mm}
\begin{figure}[t!]
   \begin{picture}(140,50)(6,-3)
      \put(0,0){ \makebox(70,50)[tl]{ \psfig{file=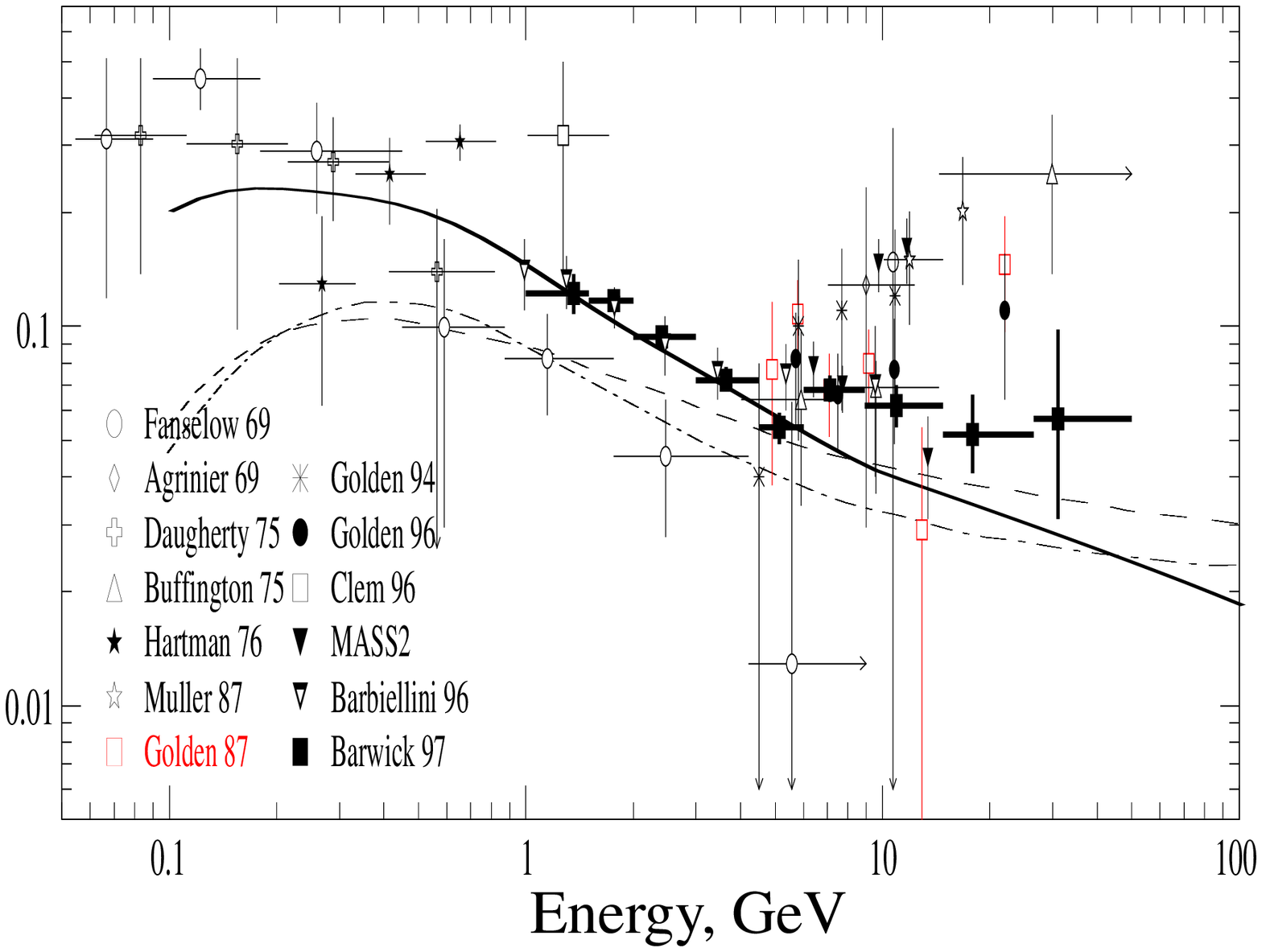,%
         height=\figheight,width=\figwidth,clip=}}}
      \put(69,0){ \makebox(70,50)[tl]{ \psfig{file=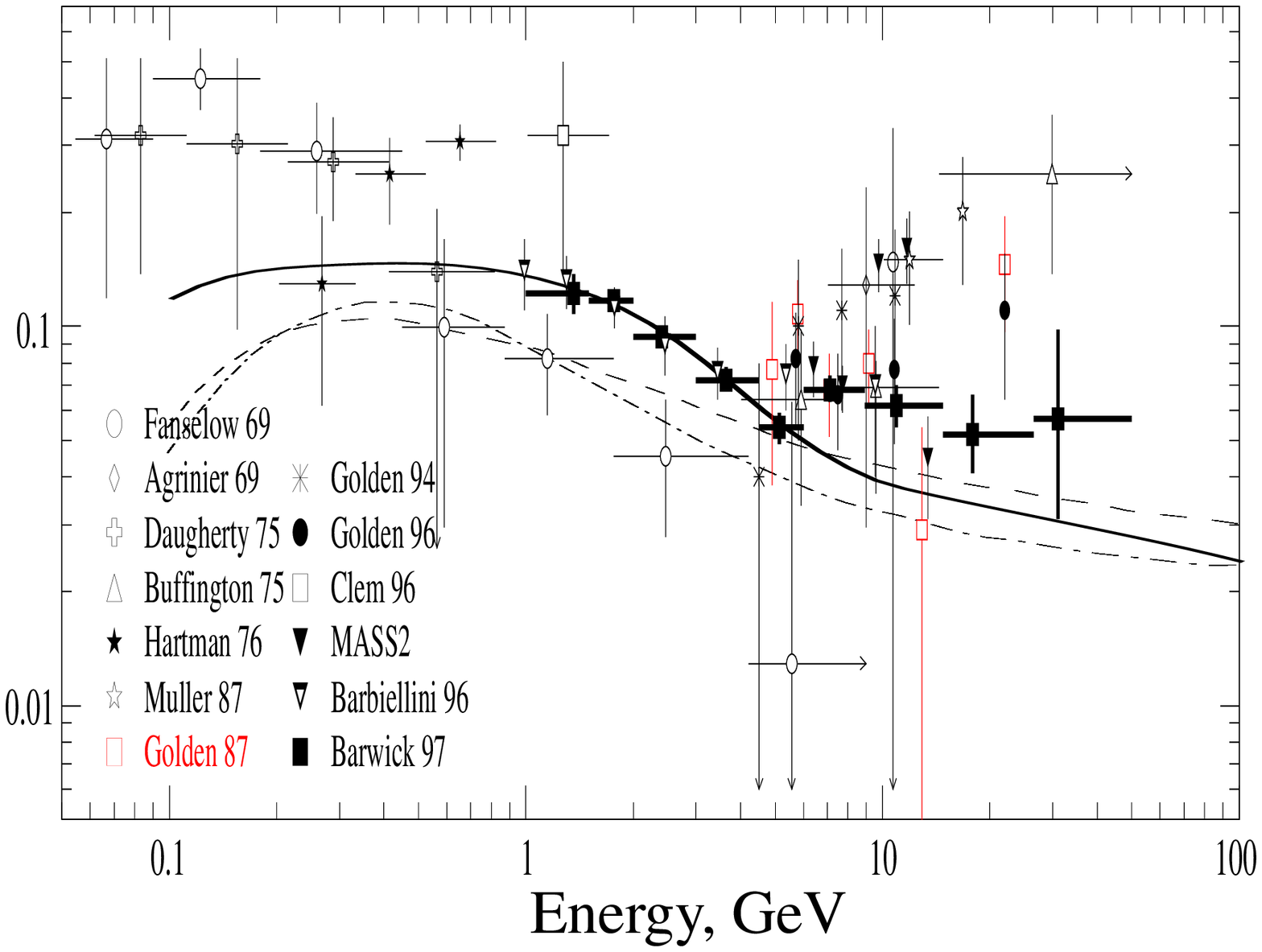,%
         height=\figheight,width=\figwidth,clip=}}}
   \end{picture}
   \caption[]{%
Positron fraction for model with no reacceleration (left panel) and
with reacceleration (right panel).  Dashed and dash-dotted lines by
Protheroe \cite{Protheroe82}: his predictions of the leaky-box and
diffusive halo models respectively.  The data collection is taken from
\cite{Barwick97}.}
   \label{fig7}
\end{figure}

\end{document}